\documentclass[twocolumn,secnumarabic,amssymb, nobibnotes, aps, prd, superscriptaddress]{revtex4-1}
\usepackage[T1]{fontenc}
\usepackage{graphicx}
\usepackage{amsmath}
\usepackage{upgreek}
\usepackage{natbib}
\usepackage{SIunits}

\begin{document}
\title{Sensing adsorption kinetics through slip velocity measurements of polymer melts}

\author{Marceau H\'enot}
\affiliation{Laboratoire de Physique des Solides, CNRS, Univ. Paris-Sud, Universit\'e
Paris-Saclay, 91405 Orsay Cedex, France}
\author{Eric Drockenmuller}
\affiliation{Univ Lyon, Universit\'e Lyon 1, CNRS, Ing\'enierie des Mat\'eriaux Polym\`eres, UMR 5223, F-69003, Lyon, France}
\author{Liliane L\'eger}
\affiliation{Laboratoire de Physique des Solides, CNRS, Univ. Paris-Sud, Universit\'e
Paris-Saclay, 91405 Orsay Cedex, France}
\author{Fr\'ed\'eric Restagno}
\email[Corresponding author: ]{frederic.restagno@u-psud.fr}
\affiliation{Laboratoire de Physique des Solides, CNRS, Univ. Paris-Sud, Universit\'e
Paris-Saclay, 91405 Orsay Cedex, France}
\date{\today}

\begin{abstract}
The evolution over time of the non-linear slip behavior of a polydimethylsiloxane (PDMS) polymer melt on a weakly adsorbing surface made of short non-entangled PDMS chains densely end-grafted to the surface of a fused silica prism has been measured. The critical shear rate at which the melt enters the nonlinear slip regime has been shown to increase with time. The adsorption kinetics of the melt on the same surface has been determined independently using ellipsometry. We show that the evolution of slip can be explained by the slow adsorption of melt chains using the Brochard-de Gennes's model.
\end{abstract}
\maketitle

\section{Introduction}
Polymer chains are known to adsorb irreversibly on some surfaces from a melt or a solution.  The conformations of adsorbed or grafted chains using neutron reflectivity has been extensively studied in particular by L. Auvray and co-workers~\cite{auvray_neutron_1986,auroy_study_1990,auroy_characterization_1991,auroy_structures_1991,auroy_local_1992,mir_density_1995} using neutron reflectivity. Neutron reflectivity also allowed to study their interactions with a solution~\citep{auroy_collapse-stretching_1992,auvray_structure_1992,auvray_irreversible_1992} or other soft matter systems~\cite{lal_perturbations_1994}. If the adsorbed polymer chains are long enough, they can entangle with the bulk chains and thus modify the mechanical coupling between the solid surface and the liquid. Their effect on liquid polymer flows is of practical importance in many fields such as polymer assisted oil recovery~\cite{de1989slip,de1989slip,chauveteau1981basic,cuenca_fluorescence_2012,cuenca_submicron_2013}, polymer extrusion~\cite{ramamurthy_1986,piau1994measurement,denn2001extrusion}, lubrication in industrial processes~\cite{cayer2008drainage} or in biological systems~\cite{dedinaite_2012}. Indeed, a fluid flowing near a wall at low Reynolds number can dissipate energy through its viscosity or through slip at the wall via friction of the last layer of fluid on the solid surface. The friction stress $\sigma$ of the liquid on the solid wall can be described using the linear response assumption made by Navier~\cite{navier} by $\sigma = k V$ where $V$ is the slip velocity and $k$ is the interfacial friction coefficient. The bulk friction coefficient $\zeta$ between two layers of molecules in the bulk can be estimated as $\zeta = \eta/a$, where $a$ is a molecular size. In most cases, particularly for simple fluids, the friction at the wall is higher than between two layers of molecules $k > \zeta$ and the slip effect is completely negligible at the macroscopic scale~\cite{navier,churaev1984slippage,pit_2000,craig_shear-dependent_2001, schmatko_2005, neto_boundary_2005}, as viscous dissipation is more favorable. However, in the case of nanoscopic scale flow~\cite{cottin-bizonne_nanorheology:_2002, secchi_massive_2016} or in the case of complex fluids, slip can play a major role. To quantify flow with slip at the wall, the slip length $b$ defined as the distance behind the wall where the velocity profile would extrapolate to zero is often used. In the case of simple shear flow, $b=V/\dot{\gamma}$ where $\dot{\gamma}$ is the shear rate.

Polymer melts on non-adsorbing surfaces can present much larger slip lengths, in the micrometric to millimetric range, so that the dissipation through slip could become of the same order, or dominate, over the viscous dissipation. This was first conjectured by de Gennes~\cite{de_gennes_1979}: he assumed that the interfacial friction of a linear polymer melt on a surface was similar to that of a simple fluid made of the same monomer, just becaus the same entities, monomers were indeed what was sliding on the surface, in both cases. As a consequence, a slip length proportional to the melt viscosity was predicted. For polymer melts submitted to shear rates in the Newtonian regime, $b$ is thus expected to be independent of the shear rate $\dot{\gamma}$. These préesdictions have been well verified experimentally, especially at high enough shear rates, on some carefully prepared ideal surfaces~\cite{wang_molecular_1997, baumchen_reduced_2009,henot_acs_2018}.

Nevertheless a large number of experimental studies reported a strong dependence of $b$ on $\dot{\gamma}$~\cite{Lim_1989, elKissi_1990,Hatzikiriakos_1992,migler_slip_1993,Drda_1995,wang_superfluid-like_1996,wang_stick-_1996,durliat_influence_1997,leger_wall_1997, massey_investigation_1998,chenneviere_direct_2016,ilton_beyond_2017,Gay1999}. For instance, the sharkskin instability that had been observed since the 40s in extrusion processes was explained by a stick slip effect resulting from nonlinear friction at the wall~\cite{rielly_1961,ramamurthy_1986,kalika_1987, denn2001extrusion}. This strong non-linearity has been attributted to the presence of adsorbed chains on the surfaces that can entangle with the melt~\cite{brochard_1992,adjari_slippage_1994,ajdari_1995,brochard_1996,jeong_2017} and undergo an entangled-desentangled transition when stretched under the effect of the local friction forces.
\begin{figure}[htbp]
  \centering
  \includegraphics[width=8.5cm]{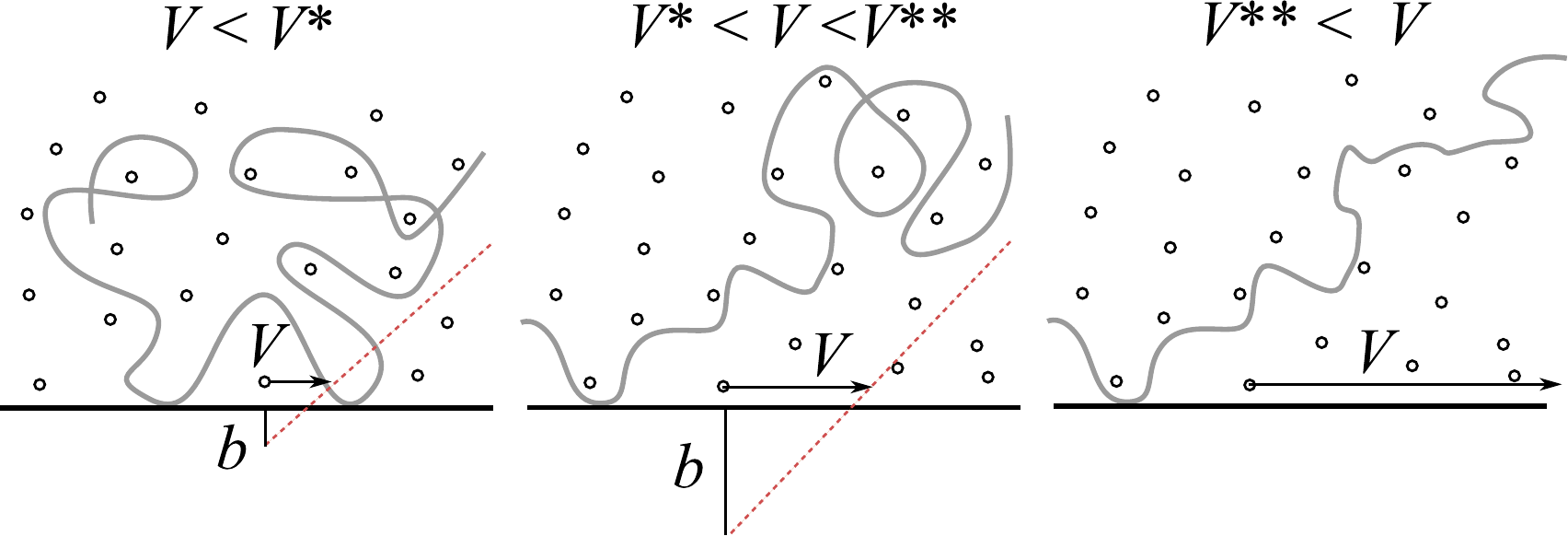}
 \caption{Schematics of the behavior of an adsorbed polymer chain when the melt is flowing. For slip velocity below $V^*$ the adsorbed chain is fully entangled with the melt leading to linear friction while when $V^*<V<V^{**}$ the tail is stretched and only the head of the chain is relaxed and entangled. This last regime is called marginal. For $V > V^{**}$, the adsorbed chain is completely stretched and fully disentangled from the melt. The friction is then low and linear.}
  \label{schema}
\end{figure}
The model developed by Brochard~\textit{et al.}~\cite{brochard_1992,adjari_slippage_1994,ajdari_1995,brochard_1996} in order to describe the effect on slip of chains end attached to the walls is illustrated in Figure~\ref{schema}. For a low enough slip velocity $V$, the attached chains are fully entangled with the melt, leading to high linear friction and small slip length. When $V$ reaches a critical value $V^*$, the attached chains start to stretch near their attachment point while their tails remain relaxed. The size of the relaxed part diminishes when the slip velocity increases leading to a progressive reduction of the number of entanglements and thus to nonlinear friction. In this regime, called the marginal regime, the shear rate stays constant at a value $\dot{\gamma}^*$. The shear stress is thus: 
\begin{equation}
\sigma^\ast =  \eta \dot{\gamma}^\ast = \nu \frac{kT}{D_\mathrm{e}}
\label{eq_brochard}
\end{equation}
where $\nu$ is the number of end attached chains per surface area, $\eta$ is the viscosity of the melt and $D_\mathrm{e}$ is the diameter of the Edward tube~\cite{deGennes_scaling}. At some sliding velocity value, the attached chains fully disentangle from the melt and the regime of ideal friction described by de Gennes is recovered. These three regimes were studied experimentally in detail by Migler, Massey and Durliat~\cite{migler_slip_1993,durliat_influence_1997,leger_wall_1997, massey_investigation_1998} for adsorbed and chemically end grafted chains. In all these previous experiments, the grafted or adsorbed layer was preprared in such a way that it was not evolving with time.

The aim of the present article is to use fluid-solid friction as a probe of the adsorption kinetic of a polymer melt on a weakly adsorbing surface. To do so, we characterized the slip behavior of a melt as a function of the melt-surface contact time in the regime where the adsorbed chains have a strong influence on the slip effect, i.e. in the marginal regime.

\section{Experiments}
The polymer fluid used was a silanol terminated PDMS melt with number average molar mass $M_{\mathrm n} = 435$~kg\(\cdot\)mol$^{-1}$ and a dispersity \DJ~$=1.05$, obtained by controlled fractionation of a commercial batch (Rh\^one-Poulenc 48V175000). The number of monomers per chain is of the order of $P\sim$~5900. For PDMS, the number of monomers within one persistence length is typically 5, hence the radius of gyration is of order 25~nm. This melt was mixed with 0.5~$\%$ by weight of fluorescently labeled photobleachable PDMS chains with a number average molar mass $M_{\mathrm n}^\star = 120$~kg\(\cdot\)mol$^{-1}$ and \DJ~$=1.17$ ($P^\ast\sim$~1600). The fluorescent chains were lab-synthesized and labeled at both ends with nitrobenzoxadiazole groups (NBD) emitting at 550~nm~\cite{leger1996,cohen_synthesis_2012} when excited at 458~nm.

The weakly adsorbing surface where the slip was measured was the polished surface of a fused silica prism, covered with end grafted short PDMS chains with an average molar mass $2$~kg$\cdot$mol$^{-1}$ ($N\sim$~27 monomers per chain), well below the average molar mass between entanglements, $M_\mathrm{e}\approx 10$~kg$\cdot$mol$^{-1}$ for PDMS~\cite{fetters_1994,leger1996}. The synthesis protocol of these end-functionalized chains along with the grafting procedure are detailed in Marzolin~\textit{et al.}~\cite{marzolin_2001}. The dry thickness of the grafted layer was $z^\ast=3.2$~nm which corresponds to a grafting density of $10^{18}$~m$^{-2}$, \textit{i.e.} 1~nm$^{-2}$. The dry thickness $z^\ast$ is higher than the Gaussian radius of gyration ($\sim 2$~nm) but smaller than the fully stretched length ($\sim 14$~nm). This corresponds to the brush regime. The advancing contact angle of water of this surface was $\theta_\mathrm{a}=112^\circ$ with an hysteresis of $5^\circ$.
The adsorption of the melt on the grafted layers was measured by rinsing the surface with toluene in a bath under agitation during 24~h. The surfaces were then dried under vacuum and the thicknesses of the remaining adsorbed layers of PDMS were measured by ellipsometry (Accurion EP3).

\begin{figure}[htbp]
  \centering
  \includegraphics[width=8.5cm]{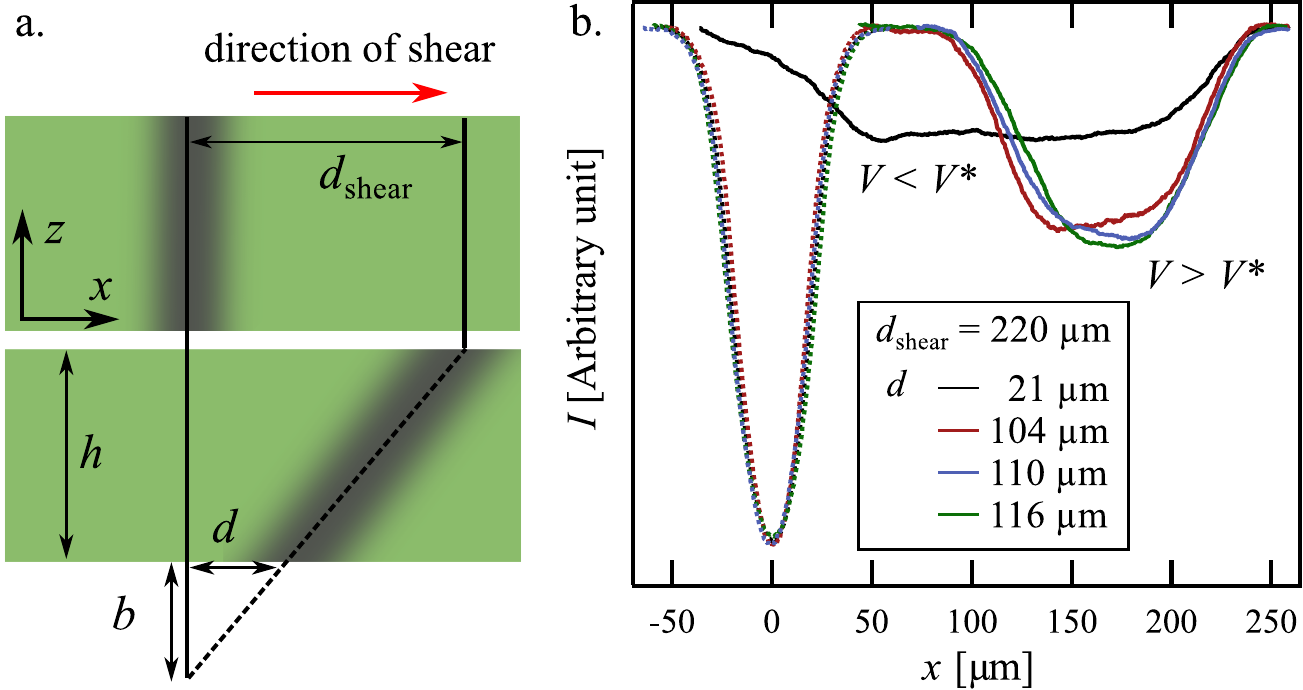}
 \caption{a. Principle of velocimetry using photobleaching. Top: schematics of the photobleached line inside the fluid. Bottom: the liquid has been sheared over a distance $d_\mathrm{shear}$, with a no slip boundary condition at the top plate while it has slipped at the bottom plate over a distance $d$ which corresponds to a slip length $b$. The black curves represent the corresponding intensity profiles integrated over $z$. b. Intensity profile of the photobleached pattern before (dotted line) and after shear (plain line) for different slip velocities below and above $V^*$. The top plate velocity goes from 425 to 965 $\mu$m$\cdot$s$^{-1}$.}
  \label{raw_data}
\end{figure}

Slip was measured using an experimental setup previously described in H\'enot~\textit{et al.}~\cite{henot_macromol_2017} relying on the analysis of the evolution under shear of the z-integrated fluorescence intensity of a pattern drawn in the sample using photobleaching. The principle of the method is illustrated in Figure~\ref{raw_data}a. The polymer melt is sandwiched between a fixed silica prism whose surface has been covered with a grafted layer of PDMS and a bare silica plate on the top. The distance $h$ between the plates is set by nickel spacers of thickness $10~\mu$m. The top plate can be displaced at a constant velocity $V_\mathrm{shear}$ over a distance $d_\mathrm{shear}$ using a stepped motor which induces a simple shear in the liquid. The distance $d_\mathrm{shear}$ is measured precisely using a linear variable differential transformer sensor. A laser spot at $\lambda= 458$~nm (Innova 90C) illuminates the sample while a home-made microscope images the fluorescence from the top. Two modes of illumination of the sample by the laser beam are available. The reading mode corresponds to a beam diameter of 2~mm when it reaches the sample. In this mode, the fluorescence can be imaged on an area of $1.5\times 2.2$~mm. The writing mode is obtained by placing a lens of focal length 10~cm just before the prism, which focuses the beam into the liquid. In this mode, the beam diameter is around $20-30~\upmu$m inside the liquid. This focused beam is used to photobleach a pattern in the fluorescent liquid by illuminating it during 800~ms with a 20~mW incident power.
Images of the fluorescence, taken before and after shear are integrated over the $y$ axis, perpendicular to the direction of the shear, in order to obtain fluorescence intensity profiles. The Gaussian profile of Eq.~\ref{eq_Ii} is fitted on the initial profile in order to determine the values of the parameters $I_0$, $A$ and $\sigma_0$.
\begin{equation}
I_\mathrm{i}(x)=I_0-A\exp \left(-\frac{x^{2}}{2\sigma^2_0}\right)
\label{eq_Ii}
\end{equation}
The profile of Eq.~\ref{eq_Is} is then fitted on the sheared profile which gives the value of the slip distance on the bottom surface $d$~\cite{henot_macromol_2017}:
\begin{equation}
I_\mathrm{s}(x)=I_0-B\left[\mathrm{erf}\left(\frac{x-d}{\sqrt{2}\sigma_0}\right)
-\mathrm{erf}\left(\frac{x-d_\mathrm{shear}}{\sqrt{2}\sigma_0}\right)\right]
\label{eq_Is}
\end{equation} with:
\begin{align*}
B=\frac{A\sigma_0\sqrt{\pi}}{2(d_\mathrm{shear}-d)}
\end{align*}
It was checked that the slip that can occur on the top plate is negligible compared to $d$ and $d_\mathrm{shear}$. This is due to the strong adsorption of PDMS on this bare silica surface.

\begin{figure}[htbp]
  \centering
  \includegraphics[width=7cm]{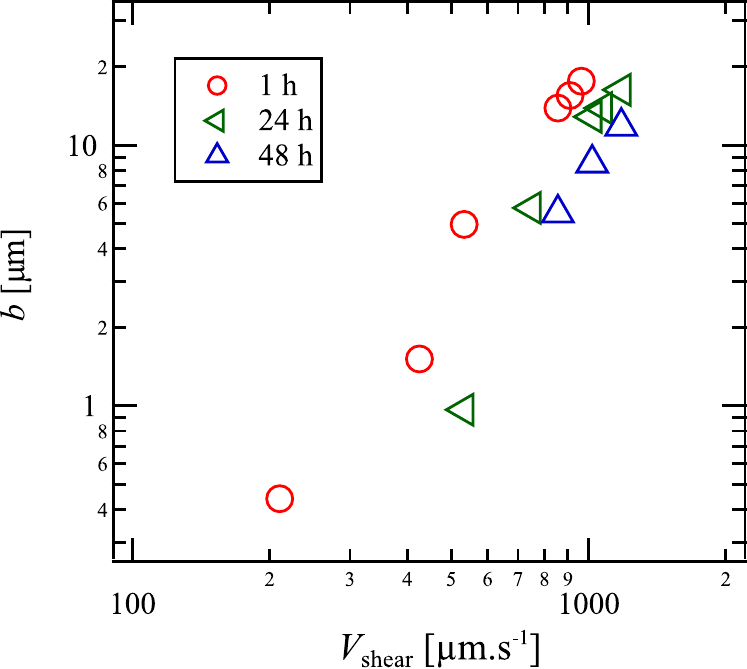}
  \caption{Slip length as a function of $V_\mathrm{shear}$ the velocity of the top plate for different times after the melt has been put in contact with the grafted layer of PDMS.}
  \label{b_Vt}
\end{figure}

Fluorescence intensity profiles before and after shear are shown in Figure~\ref{raw_data}b. They correspond to a total displacement of the top plate of $d_\mathrm{shear} = 220~\mu$m at velocities going from 425 to 965~$\mu$m$\cdot$s$^{-1}$. For low velocities the slip velocity $V$ stays low, while for higher velocities, a slip effect is well visible on the profiles. The corresponding slip lengths $b= h d /(d_\mathrm{shear}-d)$ as a function of the velocity of the top plate $V_\mathrm{shear}$ are shown in red in Figure~\ref{b_Vt}. The slip length appears to be dependent on $V_\mathrm{shear}$. These measurements were conducted within one hour after the melt had been put in contact with the freshly made weakly adsorbing surface. Same measurements performed after 24~h and 48~h are shown respectively in green and blue in Figure~\ref{b_Vt}. A measurement after 72~h was attempted but led to fracture in the melt. As can be seen, the slip behavior evolves with time. The top plate velocity required to reach a given slip length  increases with time.

\section{Discussion}
\begin{figure}[htbp]
  \centering
  \includegraphics[width=7cm]{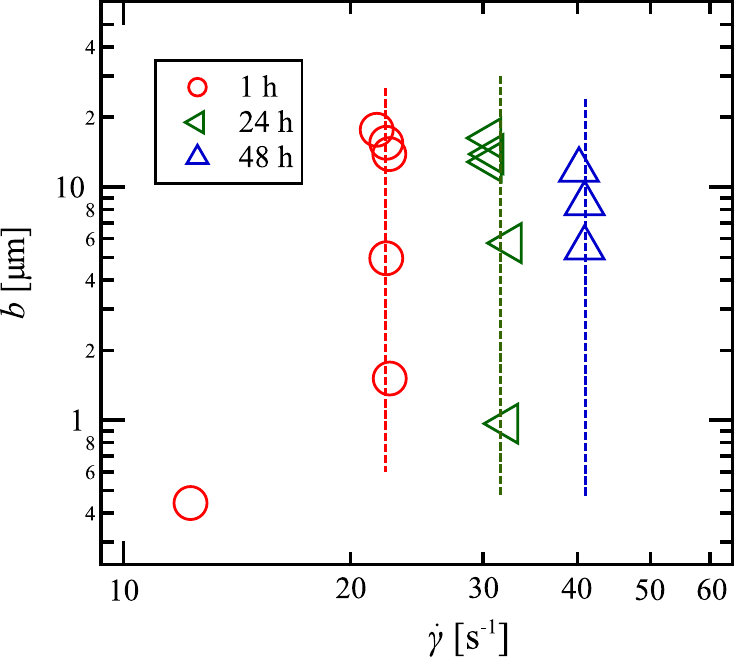}
 \caption{Slip length as a function of the shear rate experienced by the fluid for different times after the melt has been put in contact with the grafted layer of PDMS.}
  \label{b_gam}
\end{figure}
The fact that the slip behavior of the melt depends on $V_\mathrm{shear}$ means that the friction of the liquid on the surface is non-linear. Figure~\ref{b_gam} represents the slip length as a function of the shear rate experienced by the fluid $\dot{\gamma} = (V_\mathrm{shear}-V)/h$. The absolute uncertainty on the determination of $\dot{\gamma}$ is dominated by the uncertainty on the distance between the two plates which is close to 10~$\%$. However the relative uncertainty between the measurements is of the order of 0.5~s$^{-1}$ which allows one to compare them. For each time of contact, we see a regime where the slip length evolves with $V_\mathrm{shear}$, while the shear rate remains constant. This is characteristic of the so-called marginal regime described by Equ~\ref{eq_brochard} in Brochard-de Gennes's model. The point around 12~s$^{-1}$ after 1~h of contact thus corresponds to the fully entangled regime ($V<V^\ast$).

\begin{figure}[htbp]
  \centering
  \includegraphics[width=8.5cm]{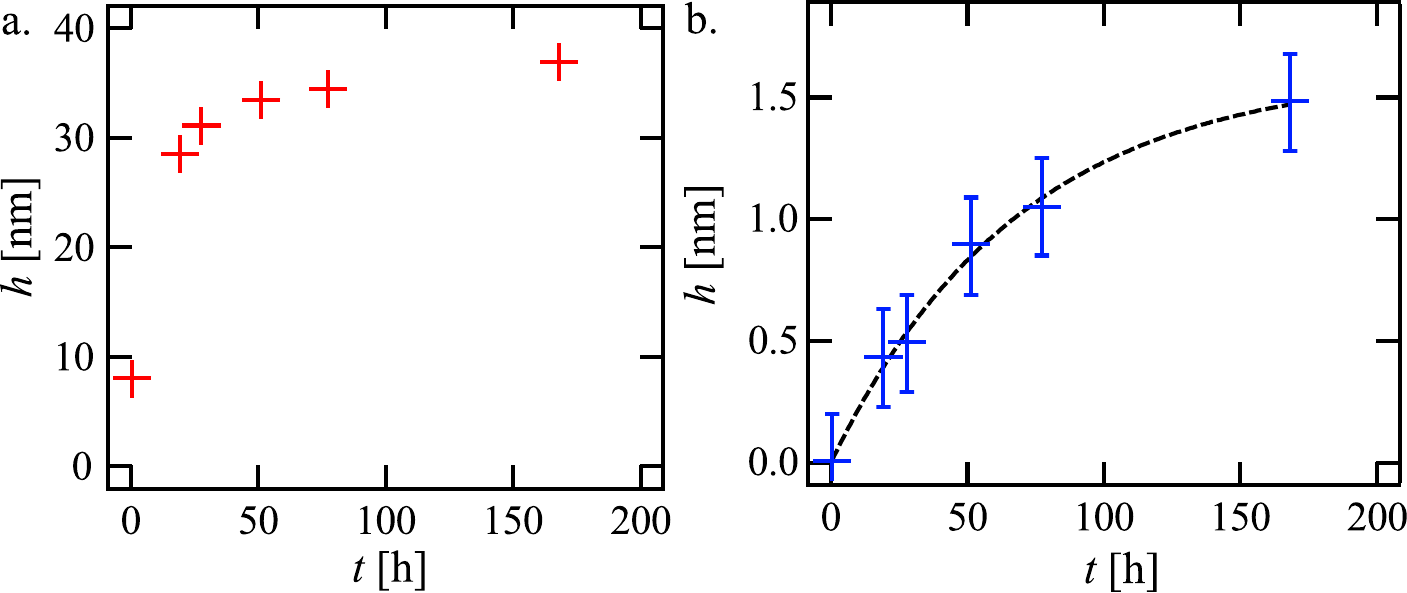}
 \caption{Thickness of the dry layer of PDMS remaining on a bare silicon wafer (a) and on the grafted layer (b) after rinsing as a function of the duration of the contact between the melt and the surface at room temperature. The back dashed curve is an exponential adjustment of the data.}
  \label{h_t}
\end{figure}

Similar weakly adsorbing Short PDMS grafted surfaces were preprared on silicon wafers, and used as tests surfaces on which the melt was deposited. These samples were kept at ambient temperature during a variable duration and then rinsed thoroughly with toluene and dried under vacuum. The dry thickness $h$ of the remaining layer of adsorbed chains was measured by ellipsometry. The same experiment was also conducted on bare silicon wafers. The results are shown in Figure~\ref{h_t}.
As expected the adsorption is strong and rapid on bare silicon wafers. This is quite different from what happens on the surface covered by a protecting layer of short grafted chains, where the free chains slowly adsorb at ambient temperature. The density of adsorbed chains is related to the dry thickness $h$ by:
\begin{equation}
\nu = h \frac{\rho \mathcal{N}_\mathrm{A}}{M_\mathrm{n}}
\label{eq_h_nu}
\end{equation}
Where $\rho = 965~$kg$\cdot$m$^{-3}$ is the density of the PDMS and $\mathcal{N}_\mathrm{A}$ is the Avogadro number. To be more precise, the maximum adsorption on the weakly adsorbing surface is twenty times lower than that on bare silicon wafer (i.e. 1.5 and 35~nm respectively). Indeed, in order to reach the underlying silicon oxyde surface, the free chain have to stretch the relatively densely grafted short chains which induces an entropy cost. However the grafted chains are short and the melt chains do adsorb, overcoming the entropy penalty or adsorbing in holes of the grafted layer. This leads to thin but measurable adsorbed layers which still have a strong effect on the slip behavior. The maximum density for the long adsorbed chains on the grafted layer is around $10^{15}$~m$^{-2}$ meaning that the molecules are separated by at least 30~nm. This is comparable to the radius of gyration. Hence, the adsorption is in the mushroom regime.

The shear rate experienced by the fluid stays constant in the marginal regime. This is consistent with early experiments done by Migler~\textit{et al.}~\cite{migler_slip_1993} and Massey~\textit{et al.}~\cite{massey_investigation_1998} with PDMS melts having higher molar masses ($M>610$~kg$\cdot$mol$^{-1}$) and corresponds well to the theoreticallypredicted  behavior. The shift with time in $\dot{\gamma}^{*}$ observed in Figure~\ref{b_gam} can be interpreted in terms of an increase of the number of adsorbed chains, and related to the evolution of the dry thickness of the adsorbed layer.
\begin{figure}[htbp]
  \centering
  \includegraphics[width=7cm]{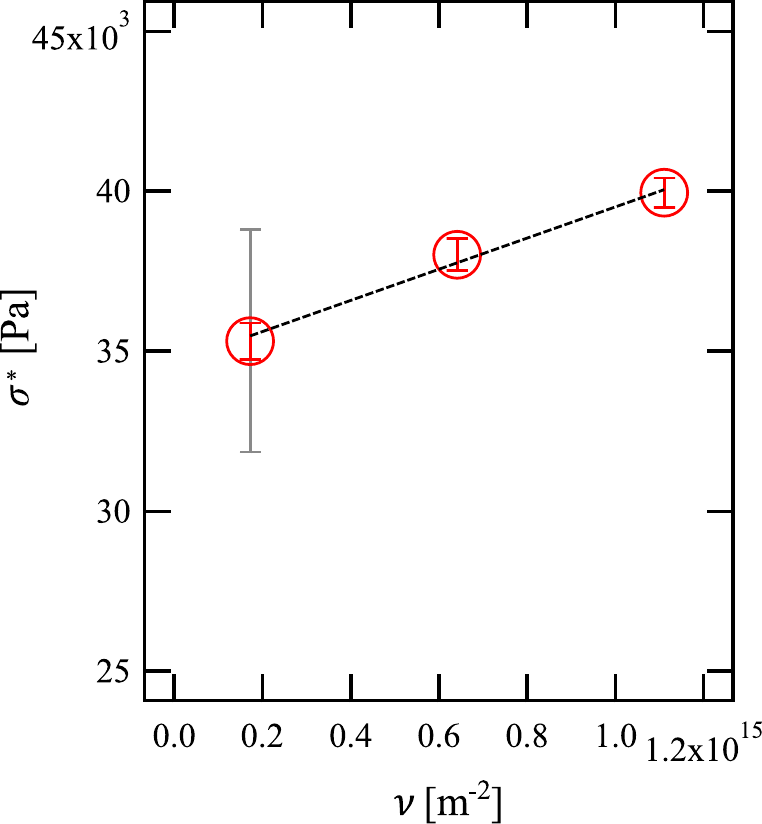}
 \caption{Product of the critical shear rate in the marginal regime and the melt viscosity at this shear rate as a function of the density of adsorbed chains. The red error bars represent the relative uncertainty between the measurements while the gray error bar represents the absolute error. The dashed black line is a linear fit.}
  \label{gam_nu}
\end{figure}
Rheological measurements performed on the melt (see supplementary materials) showed that the viscosity of the melt starts to drop around 0.1~s$^{-1}$ due to shear thinning effect. In order to take this shear thinning into account, $\sigma^\ast$ was plotted in Figure~\ref{gam_nu} as a function of $\nu$ interpolated from the measurements of Figure~\ref{h_t} and using Eq~\ref{eq_h_nu}. The relationship between these two quantities is of the form $\sigma^\ast= A\times \nu +\sigma^\ast_0$ with $A = 4.9 \pm 0.7 \times 10^{-12}$~N and $\sigma^\ast_0 = 35$~kPa. Both the slope and the intercept at origin need be discussed. 

Let us start by discussing the slope $A$. The Brochard model allows one to estimate this slope in the case of end grafted chains with no other attachment point to the surface than the grafted extremity. The value of $D_\mathrm{e} = 2.6$~nm was estimated for PDMS by Ajdari~\textit{et al.}~\cite{adjari_slippage_1994}. Using Eq.~\ref{eq_brochard}, this leads to $A_\mathrm{th}\approx 1.5\times 10^{-12}$~N. The order of magnitude is quite close to what we have obtained, but we need to account for the factor three of difference. In fact, the situation of adsorbed chains is more complex than that of end grafted chains, as adsorption on the surface can take place through any monomer. Long chains can thus entangle with the melt through any large enough tail and loop (larger than $D_\mathrm{e}$). This means that the onset of marginal regime should be related to a renormalized surface density $\nu'=\alpha \nu$ with $\alpha$ larger than one. A loop is expected to contribute as much as two tails for symmetry reasons: each half-loop exert a force toward the adsorption point and the tension should be zero between the two. Comparing our result to eq.~\ref{eq_brochard} we find $\alpha=3$. This seems plausible and means that loops also contribute to the slip transition.

The value of the shear stress in the zero adsorption limit $\sigma^\ast_0$ is more puzzling: it appears to be much larger than the contribution of the adsorbed chains. For each adsorption density, there are two contributions to the friction stress: one from the adsorbed chains entangled with the melt, and one from the monomeric friction of the melt on the short chains grafted layer (with no entanglements). A first estimate for this latter contribution can be obtained writing $\sigma_0(V) = k V$ where $k$ is the interfacial friction coefficient associated to monomer-monomer friction~\cite{de_gennes_1979}. At the onset of the marginal regime, the slip velocity is of the order of $V^\ast=40~\upmu$m$\cdot$s$^{-1}$. We have previously measured the interfacial friction coefficient for a PDMS melt sliding on grafted PDMS layers at high enough sliding velocities, so that melt and grafted chains were fully disentangled. We obtained $k\approx 1\times 10^{8}$~kg$\cdot$m$^{-1}\cdot$s$^{-1}$~\cite{henot_acs_2018}. This gives a contribution to the shear stress of order of 4~kPa at the onset of the marginal regime, much too small to explain the measured $B$ value. Such a large value of $B$ is unexpected, and quite puzzling to explain. 
A possible explanation could be that the melt penetrates the layer of short grafted chains. 
The increasing number of monomers pertaining to grafted chains around a free chains would leads to an enhanced interfacial monomeric friction compared to large velocities situation where $k$ has been measured. 
Such a penetration is certainly not total, as it would lead to higher adsorbed densities (we find weak adsorption densities, in the mushroom regime). Penetration however does exist, at least at rest, otherwise there would be no adsorption. 
To push further this argument and make it quantitative, more information on the degree of penetration of the melt inside the grafted layer, and also on the role of the slip velocity on this degree of interpenetration would be needed. Such information could be obtained through neutrons reflectivity techniques for example, but the small thicknesses involved make this task delicate.

\section{Conclusion}
We measured the evolution over time of the non-linear slip behavior of a PDMS melt using a velocimetry technique based on photobleaching on a weakly adsorbing surface made of short non-entangled PDMS chains densely end-grafted to the surface of a fused silica prism. By comparing the evolution of the critical shear rate at which the slip transition started, which increased with time, to the independently measured adsorption kinetics of the melt on the same surface, using a model developed by Brochard~\textit{et al.}~\cite{brochard_1996}, we showed that this effect is controlled by the slow adsorption of the chains and we extracted an average number of entangled loops and tails per adsorbed chains. We found a stress at zero adsorbed density, that we interpreted as monomeric friction of the melt on the grafted layer, higher than expected.

\section{Acknowledgments}
This  work  was  supported  by  ANR-ENCORE program (ANR-15-CE06-005).
We thank I. Antoniuk for technical help. 


\bibliographystyle{nar}
\bibliography{bibliographie}


\newpage

Supplementary Materials:\\

\begin{figure}[htbp]
  \centering
  \includegraphics[width=8.5cm]{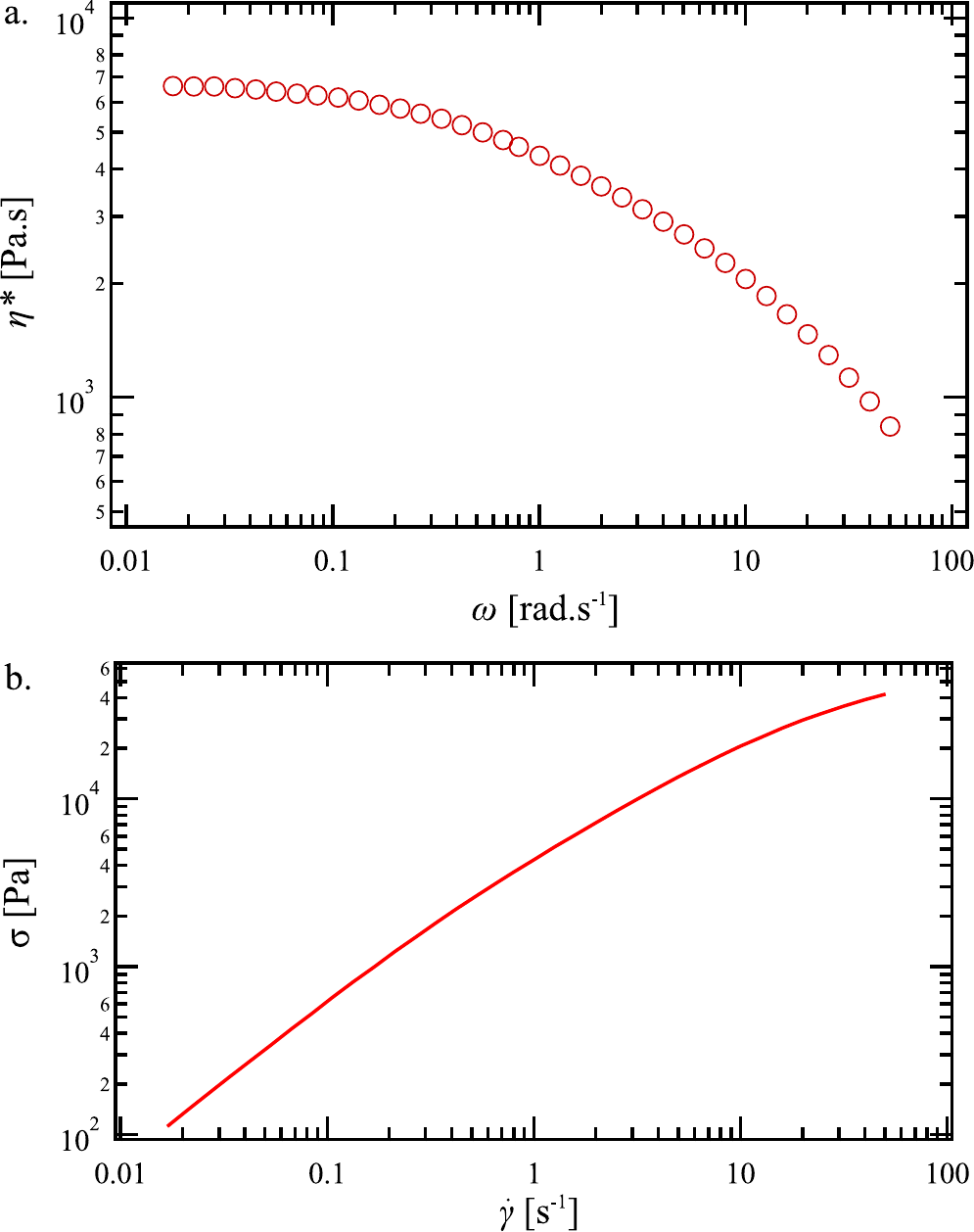}
 \caption{a. Viscosity at 20~$^{o}$C of a 435~kg$\cdot$mol$^{-1}$ PDMS melt as a function of the angular frequency. The data were measured under oscillations using a cone-plate geometry. b. Shear stress $\sigma = \eta \dot{\gamma}$ as a function of the shear rate $\dot{\gamma}$.}
  \label{rheology}
\end{figure}

Figure~\ref{rheology}.a shows the viscosity of the 435~kg$\cdot$mol$^{-1}$ PDMS melts used in this work as a function of the angular frequency. This measurement was performed at 20~$^{o}$C under oscillatory flow using a cone-plate geometry on an Anton Paar MCR302 rheometer. The relation between the viscosity obtained using oscillatory flow $\eta^{*}$ and the viscosity under shear flow $\eta$ is given by the Cox-Merx~\cite{cox1958,rheology_principles} rule $\eta^{*}(\omega) = \eta(\dot{\gamma})$. The shear stress $\sigma = \eta \dot{\gamma}$ is plotted in Figure~\ref{rheology}.b as a function of the shear rate.

\end{document}